\documentclass[pra,twocolumn,showpacs,floatfix,a4paper]{revtex4}
\usepackage{bm,color,graphicx,amsmath,txfonts,amssymb,hyperref}
\usepackage[normalem]{ulem}

\newcommand{\rp}[1]{(\ref{#1})}

\renewcommand{\thefootnote}

\newcommand{\abs}[1]{\left|{#1}\right|}

\newcommand{\av}[1]{\left\langle #1 \right\rangle}

\newcommand{\da}{^\dagger}

\newcommand{\pt}[1]{\left( #1 \right)}
\newcommand{\pq}[1]{\left[ #1 \right]}

\newcommand{\ii}{{\rm i}}

\newcommand{\nn}{{\nonumber}}

\newcommand{\GG}{{\cal G}}

\begin{document}

\title{Enhanced entanglement of two different mechanical resonators via coherent feedback}

\author{Jie Li$^{1,2,3}$, Gang Li$^{1,3}$, Stefano Zippilli$^2$, David Vitali$^2$, and Tiancai Zhang$^{1,3}$}

\affiliation{$^1$State Key Laboratory of Quantum Optics and Quantum Optics Devices, Institute of Opto-Electronics, Shanxi University, Taiyuan 030006, China \\
$^2$School of Science and Technology, Physics Division, University of Camerino, via Madonna delle Carceri, 9, I-62032 Camerino (MC), Italy, and INFN, Sezione di Perugia, Italy  \\
$^3$Collaborative Innovation Center of Extreme Optics, Shanxi University, Taiyuan 030006, China}

\begin{abstract}
It was shown [New J. Phys. 17, 103037 (2015)] that large and robust entanglement between two different mechanical resonators could be achieved, either dynamically or in the steady state, in an optomechanical system in which the two mechanical resonators are coupled to a single cavity mode driven by a suitably chosen two-tone field. An important limitation of the scheme is that the cavity decay rate must be much smaller than the two mechanical frequencies and their difference. Here we show that the entanglement can be remarkably enhanced, and the validity of the scheme can be largely extended, by adding a coherent feedback loop that effectively reduces the cavity decay rate.
\end{abstract}

\pacs{03.67.Bg, 42.50.Lc, 42.50.Wk, 85.85.+j}

\date{\today}
\maketitle

\section{Introduction}

The possibility of observing entangled states of macroscopic, massive objects is relevant to researches related, for example, to the study of the quantum-to-classical transition~\cite{Leggett,Zurek}, of wave-function collapse theories~\cite{Bassi,Bassi2,Jie} and of gravitational quantum physics~\cite{focusNJP}. However, the preparation of entanglement between massive objects is hindered by environmental noises which become hardly controllable for
large-scale systems.
Besides, it is suggested that gravitationally induced decoherence~\cite{Diosi} may also play its role in degrading the superposition and entanglement of massive objects. Recently, it has been shown that self-gravity of a macroscopic mechanical object may affect the quantum dynamics of its center-of-mass motion~\cite{Chen,Giulini,Jie2}, and, as a result, it may affect the entanglement of the motional states of two or more large objects. The ability to generate entanglement of massive objects can, therefore, be extremely useful in designing tests of these fundamental theories.

At the mesoscopic level, entanglement has been demonstrated in various systems: e.g., in two atomic ensembles~\cite{Polzik}, in two Josephson-junction qubits~\cite{Berkley,Steffen}, and in an electro-mechanical system~\cite{Lehnert}. However, entanglement between two mechanical resonators (MRs) has been demonstrated only at the microscopic level, in the case of two trapped ions~\cite{Jost}, and between two single-phonon excitations in nanodiamonds~\cite{Lee}. Optomechanics, addressing the coupling of optical and mechanical degrees of freedom via radiation pressure, provides an ideal platform to prepare quantum states in mechanical systems~\cite{OMRMP}. Many schemes, which use optomechanical and/or electromechanical systems, have been put forward for the generation of entanglement between two MRs. They exploit, for example, radiation pressure~\cite{PRL02,jopa,genesNJP,hartmann}, the transfer of entanglement~\cite{Peng03,Jie13,Ge13} and squeezing~\cite{EPL} from optical fields, conditional measurements on light modes~\cite{entswap,bjorke,mehdi1,woolley,mehdi2,Savona}, mechanical nonlinearities~\cite{Nori} and parametric drivings~\cite{Bowen}. Recently, reservoir engineering ideas~\cite{poyatos,davidovich,zoller,cirac,pielawa} have been applied to optomechanical scenarios by exploiting properly chosen multi-frequency drivings
~\cite{Clerk,Tan,schmidt,WoolleyClerk,Abdi2,Buchmann,Jie3,AnLi,Asjad} in order
to achieve robust entanglement. Similar ideas have been used to generate entangled pairs of MRs in a harmonic chain~\cite{Stefano} and cluster states of a large number of MRs~\cite{Ferraro}.

In this paper, we aim to further improve the results obtained in the scheme \cite{Jie3} by introducing a coherent feedback loop~\cite{cf1,cf2,cf3,cf4,cf5,cf6}. Differently from the conventional measurement based feedback~\cite{MF,MF1,MF2,MF3}, the non-measurement based, hence backaction free, coherent feedback shows advantages in many aspects: e.g., in cooling~\cite{Mabuchi} and suppressing noises~\cite{Yang} of MRs, in squeezing optical field~\cite{sqz0,sqz1,sqz2,sqz3,sqz4,sqz5}, in entangling optical modes~\cite{opticalEN,Xiao,Shi} and quantum networks~\cite{network}, in engineering nonlinear effects~\cite{Zhang}, and so on.
Here we apply these ideas to the preparation of entanglement between two macroscopic MRs.

The scheme~\cite{Jie3} shares similarity with the one discussed in~\cite{WoolleyClerk}. However, the former is more compact and experimental friendly in the sense that a four-tone driving is not required for the most general case of unequal optomechanical couplings~\cite{WoolleyClerk}. The protocol~\cite{Jie3} works optimally in a rotating wave approximation (RWA) regime where counter-rotating, non-resonant terms are negligible. This requires that the cavity decay rate $\kappa$ is much smaller than the two mechanical frequencies $\omega_{1,2}$ and their difference $|\omega_1-\omega_2|$, i.e., $\kappa \ll \omega_{1,2}, |\omega_1-\omega_2|$. This may limit the applicability of the scheme~\cite{Jie3} because typically the frequency of mechanical systems is not large. We show that by including coherent feedback, in which a portion of the cavity output field is returned to the input port, the effective cavity decay rate can be significantly reduced, hence relaxing the conditions of validity of the scheme, and the entanglement can be strongly enhanced.
The scheme can be optimized by controlling how much of the output light is sent back into the input port. This can be done using for example a controllable beam splitter (CBS)~\cite{sqz1,sqz2,sqz5,opticalEN}. In fact, we will show that there exist optimal values of the reflectivity of the beam splitter, of the light phase shift in the feedback loop, and of the ratio of two effective optomechanical couplings $G_{1,2}$, which yield the maximum entanglement.

The paper is organized as follows. In Section~\ref{system} we first provide the system Hamiltonian and its corresponding quantum Langevin equations (QLEs) without coherent feedback, and introduce some relevant background information already discussed in Ref.~\cite{Jie3}. We then include the feedback loop and derive the modified QLEs in Sec.~\ref{cfl}. In Section~\ref{results}, we present the results, and
finally we make our conclusions in Section~\ref{conc}.

\section{The system with no feedback
}
\label{system}

\begin{figure}[t]
\centering
\includegraphics[width=3.3in]{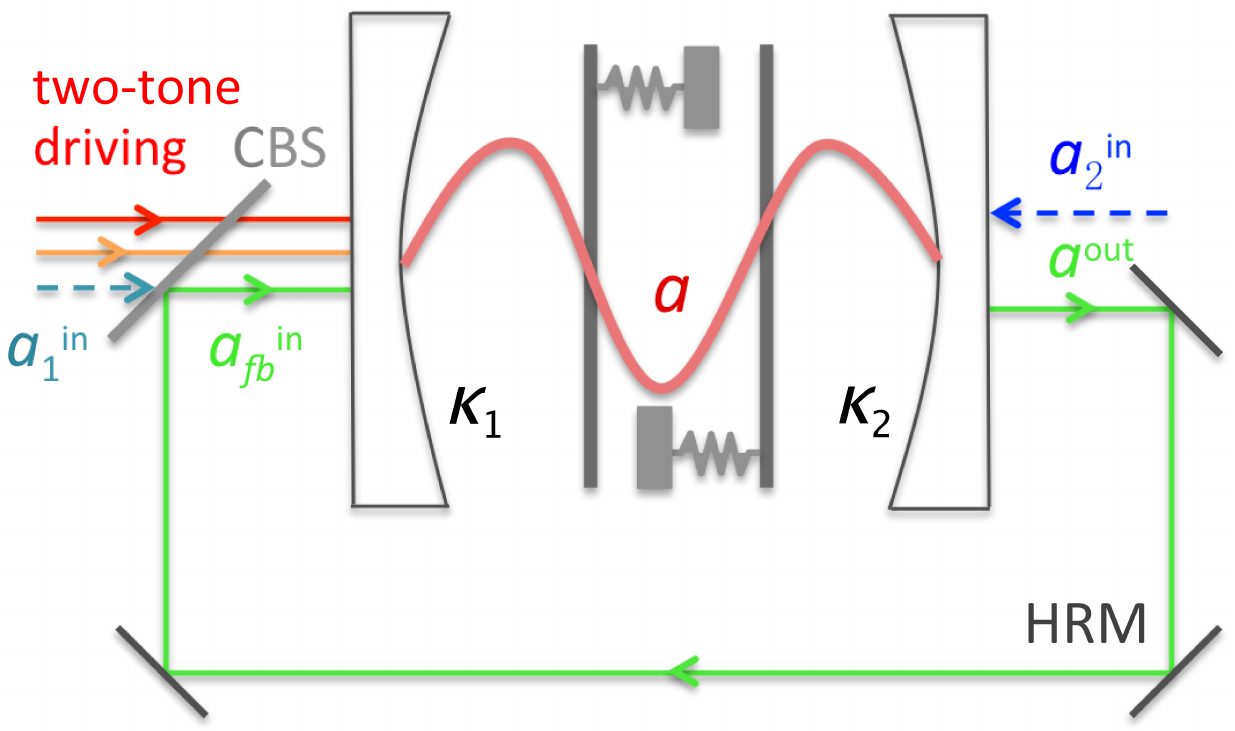}
\caption{Sketch of the system with the coherent feedback loop. The output field of the cavity is fed back into the input port through highly reflective mirrors (HRM) and a controllable beam splitter (CBS) with tunable reflection coefficient $r_B$. $a_1^{\rm in}$ and $a_2^{\rm in}$ denote vacuum noises entering into the system through the CBS and the output cavity mirror, respectively.}
\label{fig1}
\end{figure}

We study two MRs, with frequencies $\omega_1$ and $\omega_2$, which interact
with a mode of an optical cavity at frequency
$\omega_c$. The cavity is bichromatically driven at the two frequencies $\omega_{L1}=\omega_0+\omega_1$ and $\omega_{L2}=\omega_0-\omega_2$, and the reference frequency $\omega_0$ is slightly detuned from the cavity resonance by $\Delta_0=\omega_c-\omega_0$. This means that the cavity mode is simultaneously driven close to the blue sideband associated with the MR with frequency $\omega_1$, and close to the red sideband associated with the MR with frequency $\omega_2$. The system Hamiltonian, in a reference frame rotating at the frequency $\omega_0$, is given by
\begin{equation}
\begin{split}
\hat{H}&=\hbar\Delta_0  \hat{a}^\dagger \hat{a} + \hbar\sum_{j=1}^2\omega_j \hat{b}^\dagger_j \hat{b}_j + \hbar \sum_{j=1}^2 g_j  \hat{a}^\dagger \hat{a} \left(\hat{b}_j+\hat b_j^\dagger\right)  \\
&+\hbar\pq{\pt{E_1 e^{-i\omega_1t}+E_2 e^{i\omega_2t}}\hat{a}^\dagger +{\rm H.C.}},
\end{split}
\label{haml}
\end{equation}
where $\hat{a}$ and $\hat{b}_{1,2}$ are the annihilation operators of the cavity mode and the mechanical modes, respectively, $g_j$ is the single-photon optomechanical coupling to the $j$-th MR, and $E_j$ is the coupling between the driving laser and the cavity field, which is related to the pump power $P_j$ and the cavity decay rate $\kappa_1$ by $E_j=\sqrt{2 P_j \kappa_1/\hbar \omega_{Lj}}$, where $\kappa_1$, together with $\kappa_2$, are, respectively, the cavity decay rates due to the transmission through the two cavity mirrors (see Fig.~\ref{fig1}).

The system dynamics can be effectively studied by linearizing the optomechanical interaction under the assumption of sufficiently strong pump fields.
In this case, the average fields for both the cavity, $\alpha(t)$, and the mechanical degrees of freedom, $\beta_j(t)$, are large, and one can simplify the interaction Hamiltonian at the lowest order in the field fluctuations $\delta \hat a(t)=\hat a(t)-\alpha(t)$ and $\delta \hat b_j(t)=\hat b_j(t)-\beta_j(t)$
~\cite{Jie3}.
Differently from the standard approach used in the analysis of optomechanical systems~\cite{OMRMP}, here the average fields are time dependent as a result of the bichromatic driving field. However, approximated, time independent equations for the system dynamics can be derived by focusing only on the dominant resonant processes (the detailed study of this derivation and the numerical analysis of its validity can be found in Ref.~\cite{Jie3}). 
In particular, it is possible to neglect the non-resonant processes under the assumption~\cite{Jie3}
\begin{eqnarray}\label{cond01}
\abs{g_j\,\frac{E_j}{\omega_j}}  ,\,\, \kappa_{1,2}  \ll \omega_{1,2}, \, \abs{\omega_1-\omega_2} .
\label{condQLEs}
\end{eqnarray}
We remark that the condition in Eq.~\rp{cond01} requires significantly different mechanical frequencies in order to suppress specific optomechanical processes which would, otherwise, inhibit the proper operation of the scheme~\cite{Jie3}. When these conditions are fulfilled the system without feedback can be described by the following set of QLEs which, in the interaction picture with respect to the Hamiltonian $\hat{H}_0=\hbar\sum_{j=1}^2\omega_j \hat{b}^\dagger_j \hat{b}_j$, are given by~\cite{Jie3}
\begin{eqnarray}
 && \delta \dot{\hat{a}}=-(\kappa_1{+}\kappa_2{+}i\Delta)\delta \hat{a}-i G_1 \delta \hat{b}_1^\dagger -i G_2 \delta \hat{b}_2{+}\!\sum_{i=1}^2\!\!\sqrt{2\kappa_i} \hat{a}_i^{\rm in},  \label{a1st} \\
 &&\delta\dot{\hat{b}}_1 =-\frac{\gamma_1}{2} \delta \hat{b}_1  - i G_1 \delta \hat{a}^{\dagger}+\!\sqrt{\gamma_1} \hat{b}_1^{\rm in},  \label{b1st} \\
&&\delta\dot{\hat{b}}_2 =-\frac{\gamma_2}{2} \delta \hat{b}_2  - i G_2^* \delta \hat{a}+\!\sqrt{\gamma_2} \hat{b}_2^{\rm in}, \label{b2st}
\end{eqnarray}
where $\gamma_1$ and $\gamma_2$ are the damping rates of the two mechanical modes, the detuning $\Delta$ includes the optomechanical light-shift~\cite{Jie3}, $G_1$ and $G_2$ are the (generally complex) effective optomechanical couplings, given by
\begin{eqnarray}\label{G1G2}
G_1&=&\frac{g_1\,E_1}{\omega_1-\Delta+i(\kappa_1+\kappa_2)},\nn\\
G_2&=&\frac{g_2\, E_2}{-\omega_2-\Delta+i(\kappa_1+\kappa_2)},
\end{eqnarray}
and $\hat{a}_i^{\rm in}$, $\hat{b}_j^{\rm in}$ are the system input noise operators. Specifically  $\hat{a}_1^{\rm in}$ and $\hat{a}_2^{\rm in}$ are the input noise fields entering the two cavity mirrors, and $\hat{b}_j^{\rm in}$ describe the noise of the two MRs. Their non-zero correlation functions are
$\av{\hat{a}_i^{\rm in}(t)\, \hat{a}_i^{\rm in}(t')\da}=\delta(t{-}t')$, $\av{\hat{b}_j^{\rm in}(t)\, \hat{b}_j^{\rm in}(t')\da}{=}(\bar n_j{+}1)\delta(t-t')$ and $\av{\hat{b}_j^{\rm in}(t)\da\, \hat{b}_j^{\rm in}(t')}{=}\bar n_j\delta(t{-}t')$, with $\bar{n}_j=\left[\exp\left(\hbar \omega_j/k_B T\right)-1\right]^{-1}$ the mean thermal phonon number of the $j$-th MR, at the environmental temperature $T$.

These equations describe the interaction of the cavity mode with a Bogoliubov collective mode of the MRs with annihilation operator
\begin{eqnarray}\label{Bog}
\hat B&=&\frac{G_2\delta \hat{b}_2 + G_1\delta \hat{b}_1^\dagger}{{\GG}},
\end{eqnarray}
where $\GG=\sqrt{\abs{G_2}^2-\abs{G_1}^2}$ enters here as a normalization factor and is equal to the actual coupling between optical and mechanical modes~\cite{Jie3}: in fact the effective Hamiltonian corresponding to Eqs.~\rp{a1st}-\rp{b2st} can be written as $\hat{H}_{\rm eff}/\hbar=\Delta\, \delta\hat a\da \delta\hat a+\GG\pt{\hat{B}\da \delta\hat a+\hat{B}\, \delta\hat a\da}$.
The orthogonal collective mechanical mode, given by $\pt{G_2\delta \hat{b}_1 + G_1\delta \hat{b}_2^\dagger}/{{\GG}}$, remains, instead, decoupled from the cavity field.
Correspondingly these equations describe the exchange of excitations between the mechanical mode $\hat B$ and the optical cavity which are eventually lost by cavity decay, hence inducing the cooling of the Bogoliubov mode.
In particular the expression for the Bogoliubov mode is meaningful only for $G_2>G_1$. This is a sufficient condition for the stability of the system (i.e., under this condition the system approaches a steady state)~\cite{Jie3}. It corresponds to a situation in which the resonator that is driven on the red Stokes sideband (which describes the process of removing mechanical excitations) is more strongly coupled to the light with respect to the resonator that is driven on the blue anti-Stokes sideband (which instead describes the process of adding mechanical excitations). In other terms, this condition indicates that mechanical excitations have to be removed on a faster rate than they are added. Thereby the Bogoliubov collective mode can approach its vacuum state which corresponds to a two-mode squeezed state of the MRs, i.e., the MRs are prepared into an entangled state.
Note that in general a Bogoliubov transformation between two modes can be parametrized in terms of a squeezing parameter $s$ as $\hat B=\delta\hat b_2 \cosh s +\delta\hat b_1\da \sinh s$. Comparing this expression with Eq.~\rp{Bog} we observe that $s$ is determined by the relation $\tanh s=G_1/G_2$. Therefore, the squeezing parameter $s$, and hence the ratio $G_1/G_2$, determines how much the vacuum of the Bogoliubov mode is squeezed, and hence entangled,
in terms of the original modes. In particular, the vacuum of the Bogoliubov mode in the limit of equal coefficients $G_1/G_2\to 1$ corresponds to a maximally squeezed (and entangled) state of the MRs. 
The collective mechanical mode that is not coupled to the cavity mode remains, instead, in a thermal state defined by the thermal bath, and correspondingly the global state of the two MRs is a thermal squeezed state~\cite{Jie3}.

We note that in order to optimize this dynamics one needs, on the one hand, a sufficiently large difference between the couplings $G_1$ and $G_2$ (which implies a sufficiently large collective coupling $\GG$ such that the cooling is fast and efficient) and, on the other hand, almost equal couplings $G_1\simeq G_2$ such that the steady state approaches a two-mode squeezed state of the MRs with an extremely large squeezing $s$. Therefore, there exists an optimal value of the ratio $G_1/G_2$, determined by these two competing requirements, that corresponds to the maximum attainable entanglement. When, instead, $G_1=G_2$ the system displays no stationary mechanical entanglement, but significant entanglement between the MRs can still be achieved at finite times by, for example, driving the system with light pulses~\cite{Jie3}. In this case optimal entanglement is obtained at vanishing cavity decay rate, and, if also the thermal noise is negligible, the system dynamics reproduces, in an optomechanical setting, the S\o rensen--M\o lmer mechanism introduced in the context of trapped ions in Ref.~\cite{Sorensen} and extended to optical entanglement in an optomechanical system in Ref.~\cite{Kuzyk}.

\section{The coherent feedback loop}\label{cfl}

In Ref.~\cite{Jie3} the dynamics at both steady state and finite times have been extensively analyzed in the absence of feedback,  and it has been shown that strong entanglement can be achieved in both cases. Hereafter we will analyze how these dynamics are modified when a coherent feedback loop is applied.

The feedback loop sends the output field of the cavity
back into the input port as depicted in Fig.~\ref{fig1}. We shall work in the limit of instantaneous feedback, which is a very good approximation since typical mechanical frequencies are relatively small. Specifically, considering a 5 cm cavity with a 10 cm feedback loop, the delay time is $\sim10^{-10}$ s, hence the approximation of instant feedback remains valid for resonator frequencies as large as hundreds of MHz.
The output field is obtained using the standard input-output formula~\cite{Collett}
\begin{equation}
\hat{a}^{\rm out}=\sqrt{2\kappa_2} \delta\hat{a} - \hat{a}_2^{\rm in}.
\label{Cavityout}
\end{equation}
Correspondingly, the new cavity input is modeled as the superposition of the original input and the output field. In practice this is achieved by mixing the two fields in a beam splitter (see Fig.~\ref{fig1}) so that the input field modified by the feedback is
\begin{equation}
\hat{a}_{fb}^{\rm in}=r_B \, e^{i \theta}\,\hat{a}^{\rm out} +t_B \, \hat{a}_1^{\rm in},
\label{BSout}
\end{equation}
where $r_B$ and $t_B$ are the reflection and transmission coefficients, with $r_B^2+t_B^2=1$ for a beam splitter without absorption, and $\theta$ is an additional phase shift of the output field.
It should be noted that, in writing Eq.~\eqref{BSout} we have included all the possible losses (e.g. due to the transmission of the highly reflective mirrors) and phase shift (due to the reflection and propagation) of the light in the feedback loop into the reflection coefficient $r_B$ and the phase $\theta$ respectively. Specifically, here $r_B$ is interpreted as the real reflection coefficient of the beam splitter minus additional losses in the feedback loop, which means that in Eq.~\eqref{BSout} $r_B$ cannot take unity, but only approaches it, $0 \le r_B<1$.

Let us now analyze the system QLEs in the presence of coherent feedback.
The input noise operator modified by the feedback, defined in Eq.~\eqref{BSout}, can be used to replace the
bare input noise operator $\hat{a}_1^{\rm in}$ in Eq.~\eqref{a1st}. Thereby, using also Eq.~\eqref{Cavityout}, we find
the modified QLE for the cavity mode
\begin{equation}\label{NewQLE}
\delta \dot{\hat{a}}=-(\tilde\kappa+\ii\,\tilde\Delta)
\delta \hat{a}-i G_1 \delta \hat{b}_1^\dagger -i G_2 \delta \hat{b}_2
+\sqrt{2\,\tilde\kappa}\ \hat A^{\rm in},
\end{equation}
where we have introduced the effective cavity decay rate $\tilde\kappa$ and the detuning $\tilde\Delta$ which are modified by the feedback and are explicitly given by
\begin{eqnarray}\label{tildekappaDelta}
\tilde\kappa&=&\kappa_1+\kappa_2-2\sqrt{\kappa_1\,\kappa_2}\ r_B\ \cos\theta,
\nn\\
\tilde\Delta&=&\Delta-2\sqrt{\kappa_1\,\kappa_2}\ r_B\ \sin\theta\ .
\end{eqnarray}
Moreover the new input noise operator, which describes vacuum noise, is given by $\hat A^{\rm in}=\pq{(\!\! \sqrt{\kappa_2}-\!\!\sqrt{\kappa_1} e^{i \theta}\, r_B) \,\hat{a}_2^{\rm in}+\!\!\sqrt{\kappa_1}\ t_B \,\hat{a}_1^{\rm in}}/\sqrt{\tilde\kappa}$, and it is characterized by the correlation function $\av{\hat{A}^{\rm in} (t) \hat{A}^{\rm in}{}^\dag (t')}= \delta (t-t')$.
The new parameters can be either enhanced or reduced depending on the feedback phase. In particular, the cavity decay rate can be reduced down to zero when the cavity is symmetric with $\kappa_1=\kappa_2$, the reflectivity approaches unity $r_B\to 1$, and the phase is a multiple of $2\pi$. Correspondingly, for this value of the phase, the detuning remains unchanged.

We remark that Eq.~\rp{NewQLE} is valid under the conditions defined by Eq.~\rp{cond01}, and that the derivation of this equation follows the same procedure sketched in the previous section and analyzed in detail in Ref.~\cite{Jie3}. The modified detuning and decay rate, which result from the feedback loop, do not change the premises at the basis of the analysis reported in Ref.~\cite{Jie3} which, hence, remain valid also in this case. In particular, the smaller cavity decay rate, that is achievable within the feedback system, is important because it allows to extend the validity of the protocol over a wider range of parameter. In fact, the conditions expressed by Eq.~\rp{cond01} set stringent constrains due to the relatively small mechanical frequencies that typically characterize massive resonators. In practice, the condition on the optomechanical couplings $\abs{g_j\,\frac{E_j}{\omega_j}}  \ll \omega_{1,2}, \, \abs{\omega_1-\omega_2}$ can be easily fulfilled by adjusting the driving field power, while the condition on the cavity decay rates $\kappa_{1,2}  \ll \omega_{1,2}, \, \abs{\omega_1-\omega_2}$ is more difficult to be met. Therefore, a properly tailored feedback loop is of great help because one can properly adjust the value of the effective decay rate $\tilde{\kappa}$. Furthermore, the resulting entanglement is enhanced as well. In general a rough estimate of the cooling rate of the Bogoliubov mode is $\GG^2/\tilde\kappa$ (valid when $\Delta$ is negligible and $\GG<\tilde\kappa$). Hence
the same cooling rate (and roughly the same cooling efficiency) can be achieved by decreasing simultaneously the value of $\tilde\kappa$ and $\GG$. In particular, smaller $\GG$ is obtained with closer values of $G_1$ and $G_2$, that, in turn, correspond to stronger two-mode squeezing of the two MRs, and hence to stronger entanglement.

We finally remark that the complete suppression of the cavity decay rate is not in general the optimal limit for achieving maximum entanglement. The cooling of the Bogoliubov mode is effective if the cavity can dissipate the mechanical energy and this is efficient
if $\tilde \kappa$ is not smaller then $\GG$.
Correspondingly we expect that there exists an optimal value of the reflectivity $r_B$ which gives rise to maximum entanglement.

\begin{figure}[t]
\includegraphics[width=8cm]{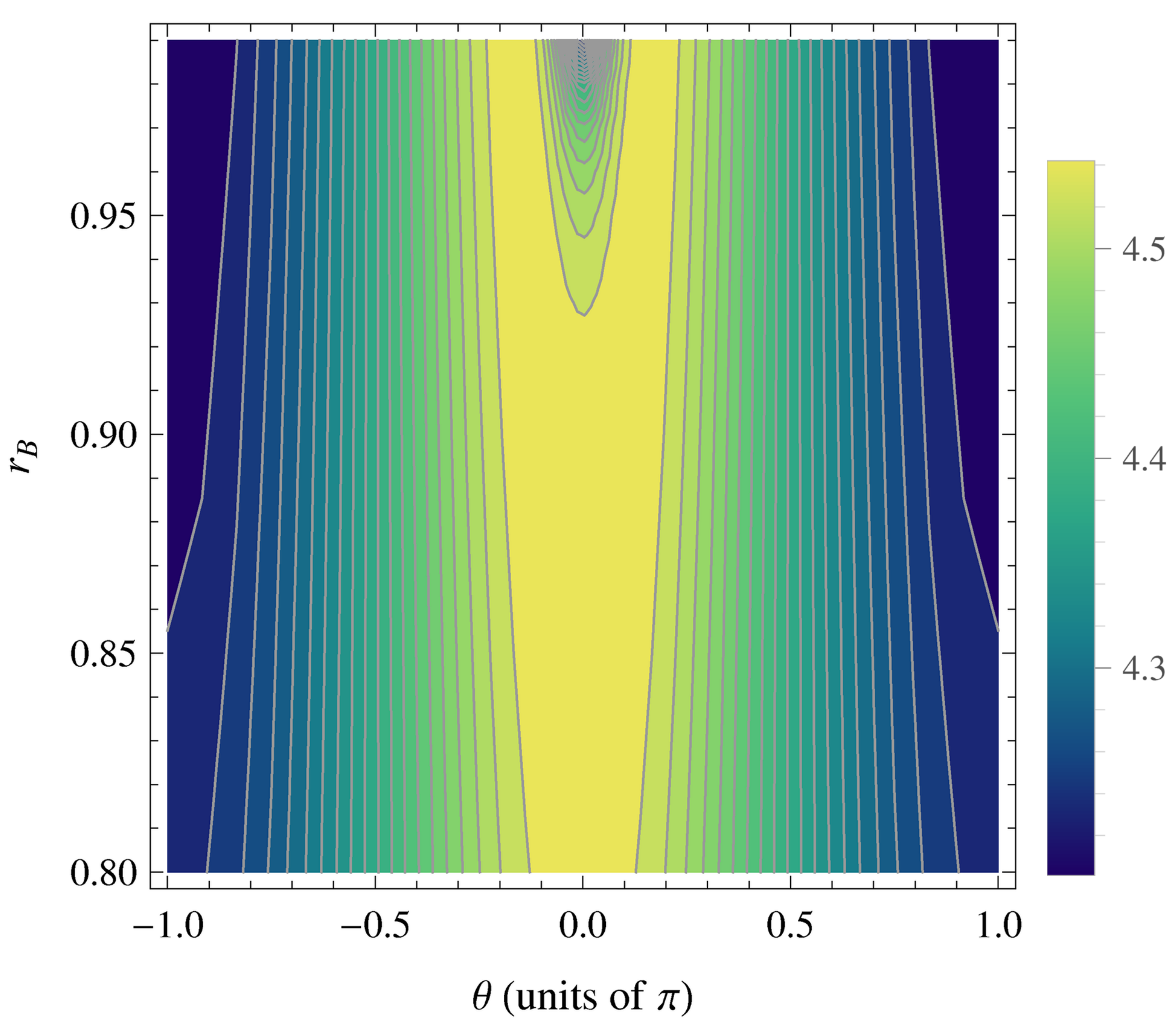}
\caption{Steady state entanglement as a function of the feedback parameters: reflectivity $r_B$ and phase shift $\theta$. The other parameters are $\bar{n}_1=\bar{n}_2=0$, $G_1=0.99 G_2$, $\gamma_1=\gamma_2=10$ Hz, $G_2=2\kappa_1=2\kappa_2=10^5$ Hz, and $\tilde\Delta = 0$.}
\label{fig2a}
\end{figure}

\begin{figure*}[t]
\hskip-0.1cm{\bf (a)}\hskip8cm{\bf (b)}\\
\hskip0cm\includegraphics[width=8cm]{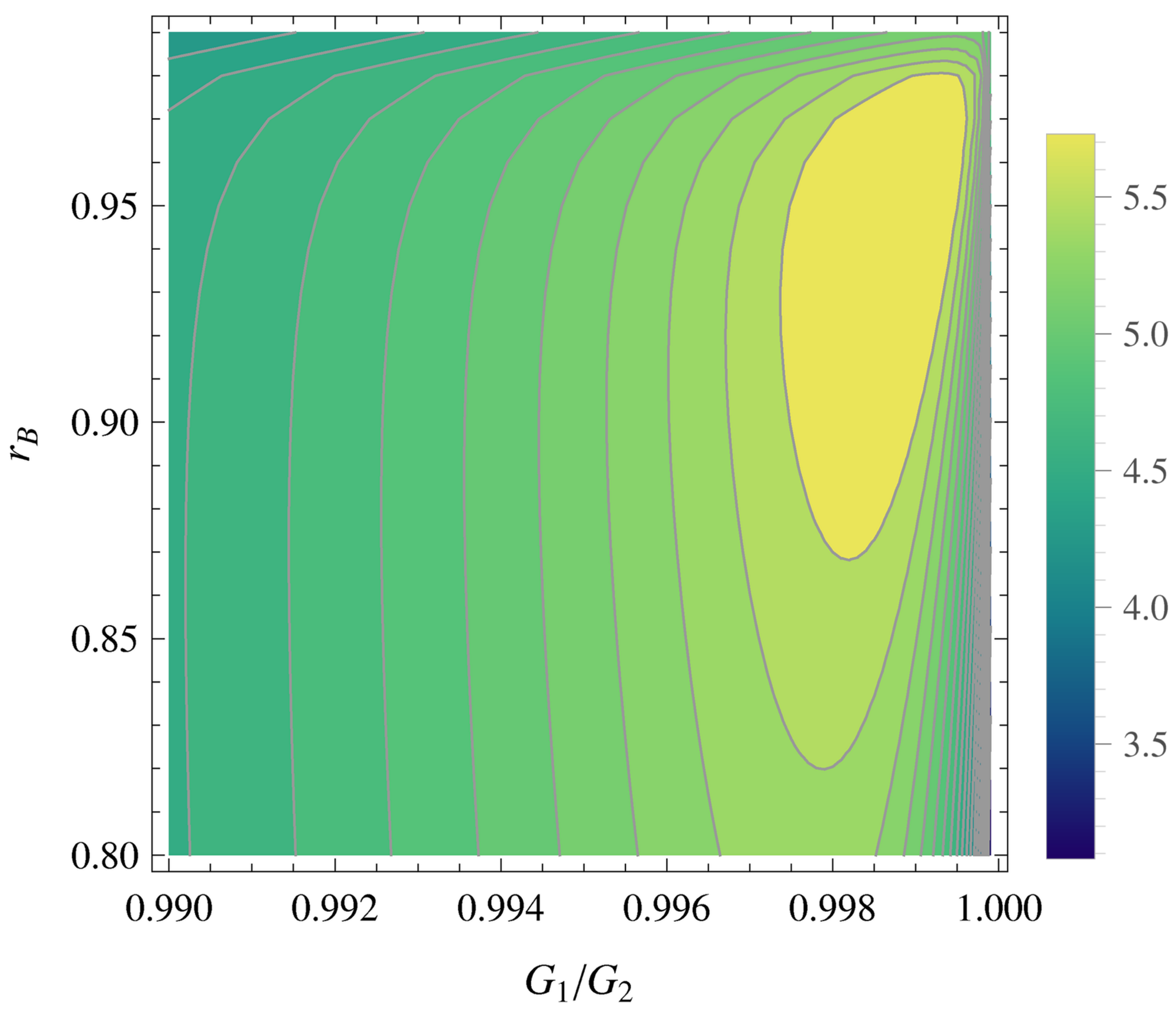}\hspace{0.4cm}
\hskip0cm\includegraphics[width=7.88cm]{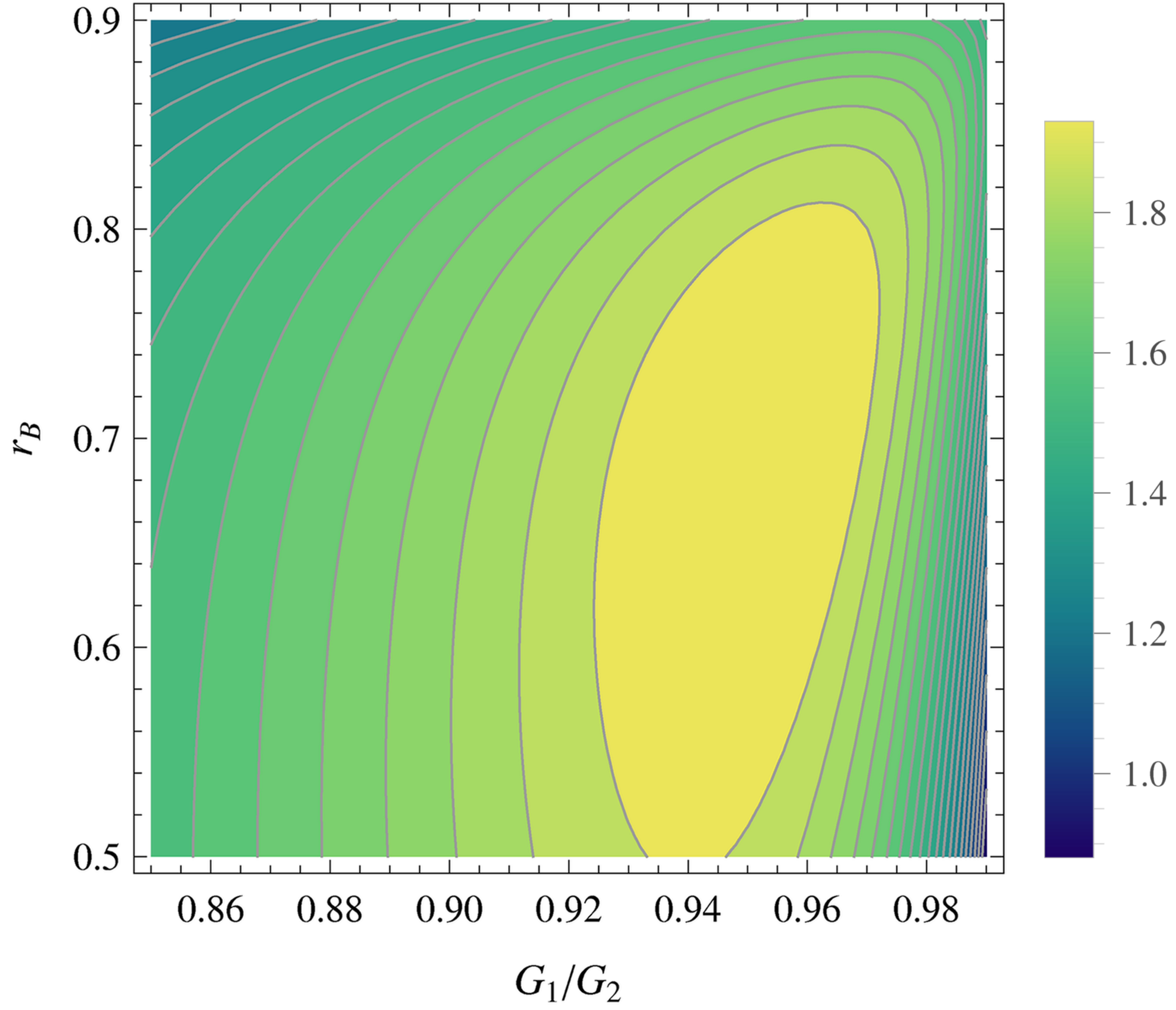}\\
\hskip-0.1cm{\bf (c)}\hskip8cm{\bf (d)}\\
\hskip-1.22cm\includegraphics[width=7.12cm]{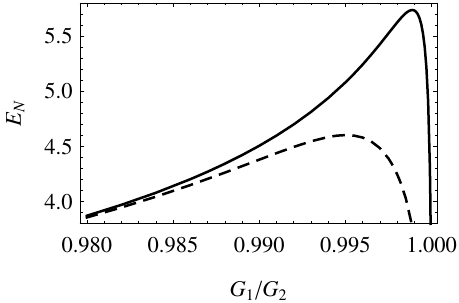}
\hskip1.2cm\includegraphics[width=6.85cm]{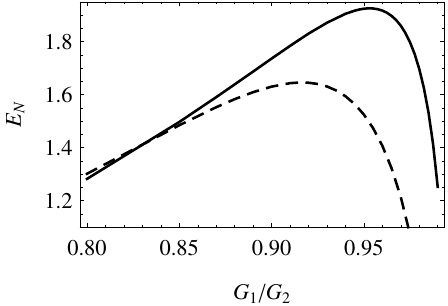}
\caption{(a), (b) Contour plot for the steady-state entanglement $E_N$ as a function of $G_1/G_2$ and $r_B$, with (a)  $\bar{n}_1=\bar{n}_2=0$, and (b) $\bar{n}_1=200$, $\bar{n}_2=100$. (c), (d) Steady-state entanglement $E_N$ as a function of $G_1/G_2$ with (solid lines) and without (dashed lines) feedback, with (c) $\bar{n}_1=\bar{n}_2=0$, $r_B=0.95$, and (d) $\bar{n}_1=200$, $\bar{n}_2=100$, $r_B=0.7$. The other parameters are $\gamma_1=\gamma_2=10$ Hz, $G_2=2\kappa=10^5$ Hz, $\Delta = 0$, and $\theta=0$.}
\label{fig2}
\end{figure*}

\section{Results}
\label{results}

In this section we report the numerical results for the entanglement, measured by the logarithmic negativity of the two MRs, evaluated by solving the system QLEs according to the procedure reported in the Appendix.

\subsection{Entanglement at steady state}

Let us first analyze the effect of the feedback on the steady state entanglement. As we have already seen the feedback modifies the value of the cavity decay rate and of the optical detuning according to the relations in Eq.~\rp{tildekappaDelta}.
We first note that, according to the discussions of Ref.~\cite{Jie3}, maximum steady state entanglement, in the absence of feedback, is achieved for $\Delta=0$. Hence, we expect to optimize the entanglement by setting the modified detuning equal to zero, namely by setting $\Delta=2\sqrt{\kappa_1\,\kappa_2}\ r_B \sin\theta$. This is the condition that we consider in this section.
Moreover, as discussed in Sec.~\ref{cfl}, we expect to observe enhanced entanglement when $\kappa$ is reduced, i.e. for $\cos\theta>0$.
This is confirmed by the results of Fig.~\ref{fig2a}. Here the system is set in a parameter regime
of robust entanglement as discussed in Ref.~\cite{Jie3}. In particular the results with no feedback correspond to the values at $\theta=\pm\pi/2$ (for which $\tilde\kappa=\kappa_1+\kappa_2$), and we observe that the entanglement increases for $\abs{\theta}<\pi/2$ under the effect of coherent feedback. We remark that maximum entanglement is obtained for a specific nonzero value of $\tilde\kappa$ that is achieved as a trade-off between two opposite needs. Specifically, on the one hand, small $\tilde\kappa$ allows for an efficient cooling of the Bogoliubov mode at smaller values of the collective coupling $\GG$, that correspond to stronger squeezing, and on the other hand, efficient cooling requires finite $\tilde\kappa\gtrsim\GG$ in order to efficiently dissipate mechanical energy.
The double peak structure at large $r_B$ is due to the fact that the optimal value of the cavity decay rate $\tilde\kappa_{\rm opt}$ is obtained for all the points along the curve in the $\theta$-$r_B$ plane that fulfill the condition $\tilde\kappa=\tilde\kappa_{\rm opt}$. We also observe that, as suggested in the previous section, optimal entanglement is not generally observed when the reflectivity is maximum $r_B\to1$, but instead, if $\theta$ is close to $2n\pi$, with $n$ integer, finite losses in the feedback (corresponding to $r_B<1$) can be instrumental to reach the optimal result.

For convenience, we fix the feedback phase at the value $\theta=2n\pi$, with $n$ integer, and study the efficiency of the scheme as a function of the system parameters. In particular, hereafter we also consider the specific situation of equal decay rates $\kappa_1=\kappa_2\equiv\kappa$ so that the effective cavity decay rate and detuning reduce to
\begin{eqnarray}\label{kD}
\tilde\kappa&=&2\kappa\pt{1-r_B},
\nn\\
\tilde\Delta&=&\Delta\ .
\end{eqnarray}

In Fig.~\ref{fig2} we study the steady state entanglement as a function of $G_1/G_2$ and $r_B$.
In particular, the contour plots in Fig.~\ref{fig2} (a) and (b) report the steady state loarithmic negativity of the two MRs as a function of both $G_1/G_2$ and $r_B$.
The solid lines in Fig.~\ref{fig2} (c) and (d) correspond, instead, to cuts of the contour plots along the value of $r_B$ that gives the maximum in (a) and (b), respectively. Finally, the dashed lines in (c) and (d) represent the results in the absence of feedback (i.e., the results at $r_B=0$). As is shown, the improvement due to the feedback is evident.
We remark that the maximum as a function of $G_1/G_2$ is found as a compromise between fast cooling of the Bogoliubov mode (large collective coupling $\GG$, possibly much larger than the mechanical dissipation rate $\sim\gamma_j\bar n_j$) and large two-mode squeezing (achieved for $G_1/G_2\to1$) corresponding to the vacuum of a Bogoliubov mode with almost equal coefficients.
The plots in (a) and (b) are computed at zero and finite temperature respectively.
As expected, the effect of the temperature is to decrease the maximum entanglement but at the same time it extends significantly the region in which the scheme performs optimally. We also note that while in the limit of zero temperature these dynamics appear quite sensitive to variations of the optomechanical couplings, at moderate temperatures, as those considered in Fig.~\ref{fig2} (b) and (d) (corresponding to tens of mK for mechanical frequency of hundreds of MHz), entanglement is much less sensitive to the specific value of the coupling ratio. 

\begin{figure*}[t]
\hskip0.9cm{\bf (a)}\hskip8.14cm{\bf (b)}\\
\hskip-0.3cm\includegraphics[width=8.3cm]{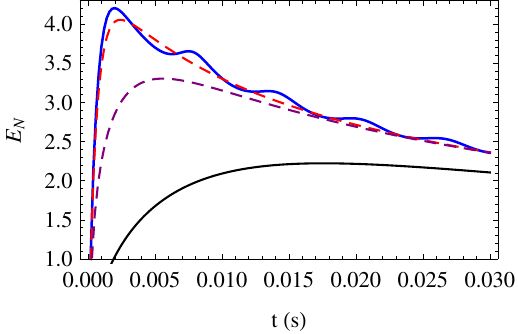}
\hspace{0.15cm}\includegraphics[width=8.3cm]{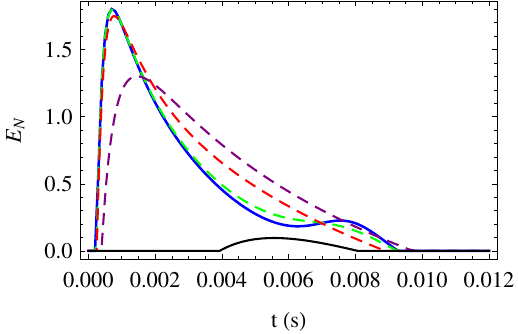}
\caption{Time evolution of the mechanical entanglement $E_N$ in the case of equal couplings, $G_1=G_2=10^4$ Hz. The evolution starts from an initial separable state with the cavity mode in the vacuum state and each MR in its thermal state. Different lines correspond to different values of $r_B$: blue line $r_B=1$ (corresponding to the perfect feedback $\tilde{\kappa}=0$), green line $r_B=0.999$, red line $r_B=0.99$, purple line $r_B=0.9$, and black line $r_B=0$ (corresponding to the case without feedback). In (a) $\bar n_1=\bar n_2=0$ (i.e. $T=0$ K), and in (b) $\bar n_1=20$ and $\bar n_2=10$.
The other parameters are $\gamma_1=\gamma_2=10$ Hz, $\kappa_1=\kappa_2=5\times10^4$ Hz, $\Delta=10^3$ Hz, and $\theta=0$.}
\label{fig3}
\end{figure*}

\subsection{Entanglement at finite time with equal couplings}

When $G_1=G_2$ no entanglement is observed at large times. In this case, however, the MRs can get entangled at finite times. This is the case studied in Fig.~\ref{fig3} which shows the time evolution of the logarithmic negativity for the two MRs. The improvement due to the feedback is even more evident in this case. Also in this case the results are evaluated with the conditions defined by Eq.~\rp{kD}, and each curve corresponds to a different value of the reflectivity $r_B$. In particular, the lower lines correspond to the case without feedback. Instead, the largest entanglement is obtained for perfect reflectivity $r_B\to 1$ which corresponds to vanishing $\tilde\kappa$, indicating that in this case, differently from the unequal couplings case studied above, it is favorable to have no losses in the feedback loop. The maximum is obtained for a relatively short interaction time, after which all the curves decay as a result of the thermal noise affecting the mechanical degrees of freedom.
The detrimental effect of the temperature $T$ is also described by the comparison of Fig.~\ref{fig3} (a) and (b) which correspond to zero and nonzero $T$, respectively. We observe that a higher temperature reduces the amount of achievable entanglement and shrinks the time window over which it is visible. Nevertheless, we observe that sizable entanglement can still be achieved in situations of very modest performance of the no-feedback scheme as shown in Fig.~\ref{fig3} (b).
We also note that in (a) the curve corresponding to perfect reflectivity (vanishing decay rate) oscillates in time. This is an evidence of the S\o rensen--M\o lmer entanglement dynamics discussed in Refs.\cite{Sorensen,Kuzyk} which, however, gradually disappears as the decay rate increases (i.e., as the reflectivity $r_B$ reduces). The S\o rensen--M\o lmer dynamics is also particularly sensitive to the thermal noise, as shown by the comparison of Fig.~\ref{fig3} (a) and (b), and a small rise of temperature washes out the oscillations of the entanglement. In order to realize these dynamics, one could consider exploiting the techniques of pulsed optomechanics~\cite{Pulsed}, by sending two weak pulsed probe beams following the detection scheme presented in~\cite{Jie3}.

\section{Conclusions}
\label{conc}

We have studied how a coherent feedback loop can be exploited to improve the efficiency of the scheme for the preparation of entanglement between two MRs reported in Ref.~\cite{Jie3}. The feedback loop
is optimized by controlling how much light is actually sent back coherently into the cavity. This can be realized using a beam splitter with tunable reflectivity. We have shown that the feedback results in a significantly reduced cavity decay rate when the reflectivity of the beam splitter is large and when the phase shift of the light in the feedback loop is properly chosen. This allows, on the one hand, to extend the validity of the original scheme, which requires a small cavity decay rate for its optimal efficiency, and on the other hand, to significantly increase the value of the entanglement as compared to the scheme without feedback. It is finally interesting to note that while, in the case of equal couplings where entanglement is generated only in the dynamical transient, maximum entanglement is obtained in the limit of a perfectly efficient feedback loop where all the output light is sent back into the cavity, the steady state entanglement is not in general optimized for perfect feedback. In fact, we have demonstrated that it can be useful to have finite losses in the feedback loop, corresponding to non perfect reflectivity $r_B$ of the beam splitter, in order to enhance the performance of the scheme.

\section*{ACKNOWLEDGMENTS}
J. L. would like to thank M. Asjad and X.-J. Jia for useful discussions and Institute of Opto-Electronics, Shanxi University, for providing a visiting position. This work has been supported by the European Commission through the H2020 FET-Proactive Project HOT, and by the Major State Basic Research Development Program of China (Grant No. 2012CB921601) and the National Natural Science Foundation of China (Grant Nos: 11634008, 11674203, 91336107, 61227902).

\section*{APPENDIX}

Here we briefly introduce the method we employ to determine the entanglement between the two MRs. The entanglement is calculated using the logarithmic negativity. It can be computed in terms of the covariance matrix of the two mechanical modes, which is obtained by solving the QLEs \eqref{b1st}, \eqref{b2st} and \eqref{NewQLE}. The QLEs can be rewritten in the following form
\begin{equation}
\dot{u} (t) = A u(t) + n(t),
\label{AppenEq}
\end{equation}
where $u$ is the vector of quadrature fluctuation operators of the two mechanical modes and one cavity mode, i.e., $u(t)=\big (\delta \hat{q}_1(t),  \delta\hat{p}_1(t),  \delta\hat{q}_2(t),  \delta\hat{p}_2(t),  \delta\hat{X}(t),  \delta\hat{Y}(t) \big )^{\rm T}$, with $\delta \hat{q}_j{=}(\delta \hat{b}_j{+}\delta \hat{b}_j^{\da})/\!\sqrt{2}$, $\delta \hat{p}_j{=} i (\delta \hat{b}_j^{\da}{-}\delta \hat{b}_j)/\!\sqrt{2}$ ($j{=}1,2$), and $\delta \hat{X}{=}(\delta \hat{a}{+}\delta \hat{a}^{\da})/\!\sqrt{2}$, $\delta \hat{Y}{=} i (\delta \hat{a}^{\da}{-}\delta \hat{a})/\!\sqrt{2}$. $A$ is the so-called drift matrix, which takes the form of
\begin{equation}
A=
\begin{pmatrix}
-\frac{\gamma_1}{2} & 0  & 0  & 0 & 0  & -G_1 \\
0 & -\frac{\gamma_1}{2}  &  0  & 0  & -G_1 & 0  \\
0 &  0  &  -\frac{\gamma_2}{2}  & 0 &  0  &  G_2 \\
0 &  0  &  0  & -\frac{\gamma_2}{2}  &  -G_2  & 0  \\
0 &  -G_1  & 0  & G_2  & -2\kappa (1{-} r_B \cos \theta)  & \tilde{\Delta} \\
-G_1 &  0  & -G_2  & 0 & -\tilde{\Delta}  & -2\kappa (1{-} r_B  \cos \theta)   \\
\end{pmatrix},
\label{drift}
\end{equation}
with the parametrs defined in the main text (we have assumed $\kappa_1=\kappa_2\equiv\kappa$). The system is stable when all the eigenvalues of the drift matrix have negative real parts, which is equivalent to the condition
\begin{equation}
    |G_2|^2 > |G_1|^2{-}\frac{\tilde\kappa\, \gamma}{2}\bigg[1+\frac{4 \tilde{\Delta}^2}{\left(\gamma+2\tilde\kappa\right)^2}\bigg],
\label{stab}
\end{equation}
for the case of equal mechanical dampings $\gamma_1=\gamma_2\equiv\gamma$~\cite{Jie3}. The term $n(t)$ is the vector of noise quadrature operators associated with the noise terms in the QLEs \eqref{b1st}, \eqref{b2st} and \eqref{NewQLE}. The formal solution of Eq. \eqref{AppenEq} is given by
\begin{equation}
u(t) = M(t) u(0) +\! \int_0^t  ds \, M(s) n(t-s),
\end{equation}
where $M(t)=e^{At}$. Therefore, the covariance matrix $V(t)$ of the system quadratures, with its entries defined as $V_{ij}=\frac{1}{2}\av{\{ u_i,u_j \} }$  ($\{\cdot,\cdot\}$ denotes an anticommutator), is obtained
\begin{equation}
V(t) = M(t) V(0) M(t)^{\rm T} +\! \int_0^t  ds \, M(s)  D M(s)^{\rm T},
\end{equation}
where $V(0)$ is the covariance matrix associated with the inital state of the system and $D$ is the diffusion matrix, whose entry is defined as
\begin{equation}
\frac{1}{2}\av{n_i(t) n_j(s)+n_j(s)n_i(t)}=D_{ij} \delta(t-s).
\end{equation}
The diffusion matrix is a diagonal matrix which, for the QLEs \eqref{b1st}, \eqref{b2st} and \eqref{NewQLE}, is $D={\rm diag}\big[ \gamma_1(\bar{n}_1+\frac{1}{2}), \gamma_1(\bar{n}_1+\frac{1}{2}), \gamma_2(\bar{n}_2+\frac{1}{2}), \gamma_2(\bar{n}_2+\frac{1}{2}), 2\kappa (1{-} r_B \cos \theta), 2\kappa (1{-} r_B \cos \theta) \big]$.

Once the covariance matrix $V(t)$ is obtained, the entanglement can then be quantified using the logarithmic negativity~\cite{Jens}:
\begin{equation}
E_N(t)=\max[0,-\ln2\tilde\nu_-(t)],
\end{equation}
where $\tilde\nu_-(t)=\min{\rm eig}|i\Omega_2\tilde{V}_m(t)|$ ($\Omega_2{=}\oplus^2_{j=1}i\sigma_y$ the so-called symplectic matrix and $\sigma_y$ the $y$-Pauli matrix) is the minimum symplectic eigenvalue of the covariance matrix $\tilde{V}_m(t)={\cal P}{V_m(t)}{\cal P}$, with $V_m(t)$ the $4\times 4$ covariance matrix related to the two mechanical modes and ${\cal P}={\rm diag}(1,1,1,-1)$ the matrix that inverts the sign of momentum of the 2nd MR, i.e., $\delta \hat{p}_2 \to -\delta \hat{p}_2$, realizing partial transposition at the level of covariance matrices~\cite{Simon}.

\end{document}